# AN ALTERNATIVE MODEL OF PARTICLE PHYSICS IN A 10-DIMENSION (PSEUDO) EUCLIDIAN SPACE-TIME

by Richard BONNEVILLE

Centre National d'Etudes Spatiales (CNES), 2 place Maurice Quentin, 75001 Paris, France

phone: +33 1 44 76 76 38, fax: +33 1 44 76 77 73, mailto: richard.bonneville@cnes.fr

## Abstract

A consistent description of the fundamental interactions of particle physics based upon the assumption of 6 real extra dimensions is presented. The usual 4-dimension space-time, a curved hypersurface with the Lorentz group as local symmetry, is assumed to be embedded in a larger flat 10-dimension space-time. Two additional assumptions are made: (i) the orthogonal 6-d space in any point of the hypersurface is invariant under the orthogonal group $\mathbf{O}(6)$ or one of its subgroups, (ii) in that space only the invariant, or distinguished, subgroups of $\mathbf{SO}(6)$ can be symmetry groups for the physical states. There are only two such possibilities, each of them corresponding to one type of particles: (i) hadrons, experiencing a gauge field associated to a real symmetry group $\mathbf{G}_{\mathrm{H}}(6)$, isomorphous to $\mathbf{SU}(3)$ and identified with the strong interaction, and (ii) leptons experiencing another gauge field associated with a real symmetry group $\mathbf{G}_{\mathrm{L}}(6)$, isomorphous to $\mathbf{SU}(2)\times\mathbf{U}(1)$ but different from the usual electro-weak coupling. Moreover both hadrons and leptons are subject to weak and electromagnetic interactions plus a scalar Higgs-type coupling. That description can be extended so as to include gravitation. Postulating a minimal Lagrangian in the full 10-dimension space-time leads to introduce one effective additional vector-type field, which does not act the same way upon hadrons and leptons, thus provoking an apparent violation of the equivalence principle.

## I. INTRODUCTION

Extra dimensions in addition to the usual 4 dimensions of space-time are a common ingredient of the theoretical attempts aiming at combining gravitation with the other fundamental interactions [1] [2] [3] [4] [5] [6]. Historically, the Kaluza – Klein approach in the 1920's assumes a $5^{th}$ dimension, which is enough to account for the tensor field of gravitation and for the vector field of electromagnetism. In addition it predicts an extra scalar field and it is a common feature of unification theories to predict the existence of additional fields and particles. However the later evidence of weak and strong interactions has shown that the original Kaluza – Klein model was not sufficient for a comprehensive description of the fundamental interactions. It is nevertheless an attractive idea to try to extend the Kaluza–Klein approach by introducing more extra dimensions so as to unify gravity with the strong and electro-weak forces. So as to be consistent with the gauge symmetries of the fundamental interactions the extra dimensions are supposed to be curled over themselves like a kind of origami whose geometry reflects the symmetries of the interactions (e.g. for electromagnetism the 1-dimension unitary group $\mathbf{U}(1)$ which is isomorphous to the 2-dimension orthogonal group $\mathbf{SO}(2)$ [N.B.: "isomorphous" means that the two groups have the same Lie algebra] or $\mathbf{SU}(3)\times\mathbf{SU}(2)\times\mathbf{U}(1)$ for the Standard Model of particle physics). It has been shown [7] that a Kaluza-Klein-type theory consistent with the usual Standard Model unifying the fundamental interactions except gravitation requires 11 dimensions. The size of the extra dimensions is generally thought to be very small, of the order of Planck's length ($10^{-35}$ m), although recent works suggest they could be much more larger, up to 0.1 mm [8]. Besides, alternative ideas for introducing additional dimensions in order to account for all the known fundamental interactions including gravitation have flourished from the 1980's, for example the string theories which require 10 dimensions.

In the present paper we present a consistent description of the fundamental interactions of particle physics within a 10-dimension space-time; it is based upon three basic assumptions on its geometry, without the need of compacting the extra dimensions. We will see what deviations from the Standard Model are resulting from this approach. In the last section, we will extend that description so as to include gravitation and we will examine the consequences.

## II. THE 10-DIMENSION SPACE-TIME

The numerous experimental confirmations of General Relativity in astrophysics lead us to acknowledge that, at least at a macroscopic scale, the geometry of space-time is that of a 4-dimension curved surface $\mathbf{\Sigma}(4)$ locally invariant under the Lorentz group $\mathbf{\Lambda}(4)$; the effect of the local curvature of the surface $\mathbf{\Sigma}(4)$ can be interpreted as resulting from a force field which is identified with gravitation [9]. Since a curved surface of dimension d can be embedded inside a flat space with (at least) $\mathrm{d}(\mathrm{d}+1)/2$ dimensions $\mathbf{\Sigma}(4)$ can be embedded inside a flat space with 10 dimensions; we will assume the following:

**Hypothesesis 1:** *our physical universe is a flat (pseudo) Euclidian space with 10 dimensions* $\mathbf{E}(10)$.

Let $\mathbf{E}_{\|}(4)_{\mathrm{M}}$ be the tangent space to $\mathbf{\Sigma}(4)$ in any point M of $\mathbf{\Sigma}(4)$, and $\mathbf{E}_{\perp}(6)_{\mathrm{M}}$ be the orthogonal space to $\mathbf{\Sigma}(4)$ in M. $\mathbf{E}_{\|}(4)_{\mathrm{M}}$ is invariant under $\mathbf{\Lambda}(4)$. $\mathbf{E}_{\perp}(6)_{\mathrm{M}}$ is a 6-dimension space. *A priori* we do not know anything about the geometry of that space but we will make a second assumption:

**Hypothesesis 2:** $\mathbf{E}_{\perp}(6)_{\mathrm{M}}$ *is supposed to be invariant under the orthogonal group* $\mathbf{O}(6)$ *or one of its subgroups.*

$\mathbf{E}_{\|}(4)_{\mathrm{M}}$ is a 4-dimension flat space in which one can define a system of 4 (pseudo) orthonormal coordinates $\{x^{\mu}\}$ with $\mu$ ranging from 0 to 3, i.e. one time coordinate $(ct)$ and 3 space coordinates $(x, y, z)$ so that the pseudo Euclidian norm $\boldsymbol{\eta}_{\mu\nu}x^{\mu}x^{\nu} = -c^2t^2 + x^2 + y^2 + z^2$ is conserved. Similarly, $\{\xi^{i}\}$ being a set of 6 orthonormal coordinates in $\mathbf{E}_{\perp}(6)_{\mathrm{M}}$, the true Euclidian norm $\sum_{i,j=1}^{6}\boldsymbol{\delta}_{ij}\xi^{i}\xi^{j} = \sum_{i=1}^{6}\left(\xi^{i}\right)^2$ is conserved.

In the 10-d space-time, the physical states are not restricted to $\mathbf{\Sigma}(4)$ but they have an extension in the other dimensions; in any point M of $\mathbf{\Sigma}(4)$ there are 4 “orbital” degrees of freedom in the tangent space $\mathbf{E}_{\|}(4)_{\mathrm{M}}$ and 6 “internal” degrees of freedom in the orthogonal space $\mathbf{E}_{\perp}(6)_{\mathrm{M}}$.

## III. HADRONS AND LEPTONS

We will henceforth consider the subgroups of $\mathbf{O}(6)$. We restrict our approach to the the special orthogonal group $\mathbf{SO}(6)$, the transformation $\xi^i \rightarrow -\xi^i$ with i ranging from 1 to 6 being equivalent in the "internal" space $\mathbf{E}_{\perp}(6)_{\mathrm{M}}$ to what parity and time reversal represent in the "orbital space" $\mathbf{E}_{\parallel}(4)_{\mathrm{M}}$. We then make a third assumption:

**Hypothesesis 3:** *only the invariant, or distinguished, subgroups of* $\mathbf{SO}(6)$ *can be symmetry groups for the physical states in* $\mathbf{E}_{\perp}(6)_{\mathrm{M}}$.

There are only two such subgroups, $\mathbf{G}_{\mathrm{L}}(6)$ and $\mathbf{G}_{\mathrm{H}}(6)$, and it can be checked that one of them, $\mathbf{G}_{\mathrm{H}}(6)$, is isomorphous to $\mathbf{SU}(3)$ and that the other one, $\mathbf{G}_{\mathrm{L}}(6)$, is isomorphous to $\mathbf{SU}(2)\times\mathbf{U}(1)$. A more detailed analysis is given in Annex A.

So only the subgoups $\mathbf{G}_{\mathrm{H}}(6)$ and $\mathbf{G}_{\mathrm{L}}(6)$ are relevant to characterize the physical states, which can accordingly be classified after the irreducible representations of $\mathbf{G}_{\mathrm{H}}(6)$ and $\mathbf{G}_{\mathrm{L}}(6)$, the same as those of $\mathbf{SU}(3)$ and $\mathbf{SU}(2)\times\mathbf{U}(1)$ respectively. We can thus evidence two types of "internal" physical states, each type being characterized by its symmetry group: the "hadronic states" characterized by the $\mathbf{G}_{\mathrm{H}}(6)$ symmetry, and the "leptonic states" characterized by the $\mathbf{G}_{\mathrm{L}}(6)$ symmetry. Those two local symmetries can be interpreted as resulting from the existence of two gauge fields respectively acting upon the hadrons and upon the leptons.

The gauge field associated to the symmetry group $\mathbf{G}_{\mathrm{H}}(6)$ is identified with the strong interaction: as $\mathbf{G}_{\mathrm{H}}(6)$ or $\mathbf{SU}(3)$ have 8 infinitesimal generators, the strong interaction is mediated through a set of 8 neutral massless vector fields, the so-called gluons; there are two fundamental conjugated 3-dimension representations, usually labelled $\{3\}$ and $\{3^*\}$, of $\mathbf{G}_{\mathrm{H}}(6)$ or $\mathbf{SU}(3)$, from which all the others representations can be built, and which can can be associated to 2 triplets of particles, the so-called quarks, hence there are 2 distinct triplets of such particles (plus their images by charge conjugation).

The gauge field associated to the symmetry group $\mathbf{G}_{\mathrm{L}}(6)$ is what we call "pseudo electro-weak interaction": it has the same symmetry as but it is intrinsically different from the usual electro-weak interaction. As $\mathbf{G}_{\mathrm{L}}(6)$ or $\mathbf{SU}(2)\times\mathbf{U}(1)$ have 4 infinitesimal generators, the “pseudo electro-weak interaction” is mediated through a set of 4 neutral massless vector fields that we will call “gluinos”. Those gluinos, which are bosons, must not be confused with the so-called gluinos of the super symmetric theories, where they are the fermionic partners of the gluons. The existence of the pseudo electro-weak interaction mediated by 4 extra bosons is the major difference with the Standard Model. There are two 2-dimension conjugated representations $\{J=1/2,\ Z=\pm 1\}$ and one 2-dimension self-conjugated representation $\{J=1/2,\ Z=0\}$ of $\mathbf{G}_{L}(6)$ or $\mathbf{SU}(2)\times\mathbf{U}(1)$, from which all the others representations can be built, and which can be associated to 3 doublets of particles, hence there are 3 distinct pairs of such particles (plus their images by charge conjugation).

## IV. ONE STEP BEYOND

### IV.1. Weak interaction

We note that $\mathbf{G}_{L}(6)$ and $\mathbf{G}_{H}(6)$ have in common $\mathbf{SO}(3)\otimes\mathbf{I}(2)$ as a maximal subgroup, where $\mathbf{I}(2)$ denotes the 2-dimension identity. The local symmetry $\mathbf{SO}(3)$, isomorphous to the special unitary symmetry $\mathbf{SU}(2)$, can be interpreted as resulting from the existence of a gauge field acting on both hadrons and leptons, which is identified with the weak interaction. It may lift the degeneracy of the multiplets associated to the irreducible representations of $\mathbf{G}_{H}(6)$ and $\mathbf{G}_{L}(6)$. Since $\mathbf{SO}(3)$ or $\mathbf{SU}(2)$ has three infinitesimal generators, the weak interaction is mediated through a triplet of vector fields, the so-called $W^{+}$, $W^{-}$ and $Z^{0}$ bosons.

### IV.2. Electromagnetic and Higgs interactions

$\mathbf{SO}(3)$ has two orthogonal subgroups: $\mathbf{SO}(2)$ isomorphous to the unitary symmetry group $\mathbf{U}(1)$ and $\mathbf{SO}(1)$ i.e. the trivial identity $\mathbf{I}(1)$. The $\mathbf{SO}(2)$ symmetry can be interpreted as resulting from the existence of a gauge field experienced by both hadrons and leptons which is identified with the electromagnetic interaction. It eventually lifts the residual degeneracy of the multiplets associated to the irreducible representations of $\mathbf{SO}(3)$. The interaction is mediated through a single massless vector field which is identified with the photon. The $\mathbf{I}(1)$ symmetry can be interpreted as featuring an additional scalar field. As a consequence, both hadrons and leptons can experience a same scalar interaction mediated through a particle which can be identified with the Higgs boson.

As $\mathbf{SO}(2)$ and $\mathbf{I}(1)$ are orthogonal subgroups of $\mathbf{SO}(3)$, the Higgs boson has no electric charge and the photon has no mass; conversely the $\mathrm{W}^{+}$, $\mathrm{W}^{-}$ and $\mathrm{Z}^{0}$ bosons have both an electric charge and a mass.

### IV.3. Geometrical interpretation

In the present approach, the $\mathbf{SO}(6)$ symmetry of $\mathbf{E}_{\perp}(6)_{\mathrm{M}}$ is spontaneously broken down to $\mathbf{G}_{\mathrm{L}}(6)$ or $\mathbf{G}_{\mathrm{H}}(6)$, which in turn are broken down to their common sub-symmetry $\mathbf{SO}(3)$, which finally is broken down to $\mathbf{SO}(2)$ or $\mathbf{I}(1)$. In geometrical terms it is possible to inscribe inside the "internal" Euclidian space $\mathbf{E}_{\perp}(6)_{\mathrm{M}}$ a 4-dimension hypersurface invariant either under $\mathbf{G}_{\mathrm{H}}(6)$ or $\mathbf{G}_{\mathrm{L}}(6)$. In any point of that hypersurface there is a tangent 3-dimensional flat space and an orthogonal 3-dimensional flat space, both being isotropic (rotational invariance $\mathbf{SO}(3)$). In that orthogonal space it is possible to inscribe a sphere; in any point of the later there is a tangent plane invariant under plane rotation i.e. $\mathbf{SO}(2)$ and an orthogonal line only invariant under identity i.e. $\mathbf{I}(1)$.

Besides, the charge conjugation simply appears as the equivalent in the "internal space" $\mathbf{E}_{\perp}(6)_{\mathrm{M}}$ of what are parity and time reversal in the "orbital space" $\mathbf{E}_{\parallel}(4)_{\mathrm{M}}$, i.e. the transformation $\xi^{i} \rightarrow -\xi^{i}$ with i ranging from 1 to 6 is analogous to $x^{\mu} \rightarrow -x^{\mu}$ with $\mu$ ranging from 0 to 3. In addition, the vector character of the gluons, gluinos, $\mathrm{W}^{+}$, $\mathrm{W}^{-}$, $\mathrm{Z}^{0}$ and photon can easily be verified (Annex B).

## V. THE GRAVITATION FIELD

In the previous sections we have separately considered the "orbital" space $\boldsymbol{\Sigma}(4)$, and the "internal" space $\mathbf{E}_{\perp}(6)_{\mathrm{M}}$ defined in any point M of $\boldsymbol{\Sigma}(4)$. $\boldsymbol{\Sigma}(4)$ is a 4-dimension space locally invariant under the Lorentz group $\boldsymbol{\Lambda}(4)$ and $\mathbf{E}_{\perp}(6)_{\mathrm{M}}$ is a 6-dimension space invariant under the distinguished subgroups $\mathbf{G}_{\mathrm{H}}(6)$ and $\mathbf{G}_{\mathrm{L}}(6)$ of $\mathbf{SO}(6)$. We now consider the full 10-dimensional space $\mathbf{E}(10)$ which is the tensor product of $\boldsymbol{\Sigma}(4)$ and $\mathbf{E}_{\perp}(6)_{\mathrm{M}}$.

The symmetry propreties of $\mathbf{E}_{\perp}(6)_{\mathrm{M}}$ have evidenced several gauge fields associated to fundamental interactions. Now gravitation can also be considered as a gauge field by which the global Lorentz invariance of special relativity is changed into a local invariance [10]. It means that on the 4-dimension hypersurface $\boldsymbol{\Sigma}(4)$ the local derivation $\partial_{\mu} = \dfrac{\partial}{\partial x^{\mu}}$ is changed into

$$\mathrm{D}_{\mu} = \partial_{\mu} + \mathbf{G}_{\mu}^{\nu}\partial_{\nu} \tag{1}$$

with $\mu$ ranging from 0 to 3, where $\mathbf{G}_{\mu}^{\nu}$ is a tensor quantity featuring the local geometry of $\boldsymbol{\Sigma}(4)$. The impulsion

$$\mathbf{p}^{\mu} = \mathrm{i}^{-1}\partial^{\mu} = \mathrm{i}^{-1}\boldsymbol{\eta}^{\mu\nu}\partial_{\nu} = \left(\mathrm{i}\frac{\partial}{\partial t}, \mathrm{i}^{-1}\vec{\nabla}\right) \tag{2}$$

is thus changed into

$$\mathbf{P}^{\mu} = \mathrm{i}^{-1}D^{\mu} = \mathbf{p}^{\mu} + \mathbf{G}_{\nu}^{\mu}\mathbf{p}^{\nu}. \tag{3}$$

There is some flexibility in the determination of the $\mathbf{G}_{\mu}^{\nu}$ 's which allows to impose the 4 conditions

$$\partial_{\mu}\mathbf{G}_{\nu}^{\mu} = 0. \tag{4}$$

Now the 4-dimension surface $\boldsymbol{\Sigma}(4)$ is supposed to be embedded in the full 10-dimension flat space-time $\mathbf{E}(10)$. In the previous section, we had separately assumed the conservation of the pseudo Euclidian norm $\boldsymbol{\eta}_{\mu\nu}x^{\mu}x^{\nu} = -c^2t^2 + x^2 + y^2 + z^2$ within $\mathbf{E}_{\parallel}(4)_{\mathrm{M}}$ and the conservation of a true Euclidian norm $\sum_{i=1}^{6}\left(\xi^{i}\right)^2$ within $\mathbf{E}_{\perp}(6)_{\mathrm{M}}$. We now postulate that $\mathbf{E}(10)$ is assumed to be a pseudo Euclidian space which preserves a pseudo norm $\boldsymbol{\eta}_{\alpha\beta}X^{\alpha}X^{\beta}$ with $\alpha$ and $\beta$ ranging from 0 to 9, i.e. $\boldsymbol{\eta}_{00} = -1$, $\boldsymbol{\eta}_{\alpha\alpha} = 1$ if $\alpha \neq 0$, $\boldsymbol{\eta}_{\alpha\beta} = 0$ otherwise. We further postulate that in a 10-dimension reference frame of $\mathbf{E}(10)$, the Lagrangian $\mathfrak{L}$ attached to a field $\Theta(x^{\alpha})$ simply is

$$\mathfrak{L} = \boldsymbol{\eta}^{\alpha\beta}\dot{\Theta}_{\alpha}^{\dagger}\dot{\Theta}_{\beta}, \text{ with } \dot{\Theta}_{\alpha} = \partial_{\alpha}\Theta \tag{5}$$

We note that the symmetry goup of $\mathbf{E}(10)$ has $\mathbf{\Lambda}(4)$ and $\mathbf{SO}(6)$ as invariant subgroups. Now taking into account the local character of those symmetries in the pseudo euclidian reference frame { $ct, x, y, z, \xi^1, \xi^2, \xi^3, \xi^4, \xi^5, \xi^6$ } attached to the point M of $\mathbf{\Sigma}(4)$, we extend to $\mathbf{E}(10)$ what has been done above with the Lorentz invariance in the case of a 4-dimension space-time. The transformation equ.(1) readily becomes

$$\partial_\alpha \to D_\alpha = \partial_\alpha + \mathbf{G}_\alpha^\beta \partial_\beta \tag{6}$$

and the Lagrangian equ.(5) is changed into

$$\mathfrak{L} = \boldsymbol{\eta}_{\alpha\beta}[\mathbf{P}^\alpha \Theta]^\dagger [\mathbf{P}^\beta \Theta] \tag{7}$$

or

$$\mathfrak{L} = \mathbf{g}_{\alpha\beta}[\mathbf{p}^\alpha \Theta]^\dagger [\mathbf{p}^\beta \Theta] \tag{8}$$

where we have introduced the effective 10x10 metric tensor

$$\mathbf{g}_{\alpha\beta} = \boldsymbol{\eta}_{\alpha\beta} + \mathbf{h}_{\alpha\beta} \tag{9}$$

with

$$\mathbf{h}_{\alpha\beta} = \boldsymbol{\eta}_{\rho\sigma}\mathbf{G}_\alpha^\rho \mathbf{G}_\beta^\sigma + \boldsymbol{\eta}_{\alpha\rho}\mathbf{G}_\beta^\rho + \boldsymbol{\eta}_{\beta\sigma}\mathbf{G}_\alpha^\sigma \tag{10}$$

[N.B.: In the present work for sake of formal simplicity we only consider transformations between (pseudo) orthonormal reference frames.]

From the expression of $\mathbf{g}^{\alpha\beta}$ as a function of the $\mathbf{G}_{\alpha\beta}$ s it is obvious that $\mathbf{g}^{\alpha\beta} = \mathbf{g}^{\beta\alpha}$. Applying the Lagrange equations to the $\boldsymbol{\Theta}$ field, i.e.

$$\partial_\alpha \frac{\partial \mathfrak{L}}{\partial \dot{\Theta}_\alpha} = \frac{\partial \mathfrak{L}}{\partial \Theta}, \tag{11}$$

gives

$$\mathbf{g}^{\alpha\beta}\partial_\alpha \partial_\beta \boldsymbol{\Theta} + (\partial_\alpha \mathbf{g}^{\alpha\beta})(\partial_\beta \boldsymbol{\Theta}) = 0 \tag{12}$$

The gauge properties give some flexibility in the determination of the $\mathbf{G}_\beta^\alpha$ 's so that we can impose the 10 conditions

$$\partial_\alpha \mathbf{G}_\beta^\alpha = 0 \tag{13}$$

from which we derive

$$\partial_\alpha \mathbf{g}^{\alpha\beta} = 0 \tag{14}$$

The equation of evolution equ.(12) thus becomes

$$\mathbf{g}^{\alpha\beta}\partial_\alpha\partial_\beta \mathbf{\Theta} = 0 \tag{15}$$

That expression means that the only accessible physical states are those whose measure is null in $\mathbf{E}(10)$. With the correspondence $\mathbf{p}^\alpha \leftrightarrow \mathrm{i}^{-1}\hbar\partial^\alpha$, one equivalently gets

$$\mathbf{g}_{\alpha\beta}\mathbf{p}^\alpha\mathbf{p}^\beta\Theta = 0. \tag{16}$$

In the local reference frame { $ct, x, y, z, \xi^1, \xi^2, \xi^3, \xi^4, \xi^5, \xi^6$ } $\mathbf{g}$ has the following form

$$\mathbf{g}^{\alpha\beta} = \left(\begin{array}{c|c} \begin{matrix}\mathbf{g}^{\mu\nu} \\ 4\times 4\end{matrix} & \begin{matrix}\mathbf{a}^{\mu j} \\ 4\times 6\end{matrix} \\ \hline \begin{matrix}\mathbf{a}^{i\nu} \\ 6\times 4\end{matrix} & \begin{matrix}\mathbf{s}^{ij} \\ 6\times 6\end{matrix} \end{array}\right) \tag{17}$$

Since $\mathbf{g}^{\alpha\beta} = \mathbf{g}^{\beta\alpha}$ and $\partial_\alpha \mathbf{g}^{\alpha\beta} = 0$, $\mathbf{g}$ has 55 distinct components, of which 45 are independent. In the previous sections where gravitation and the other fundamental interactions were disconnected, we had $\mathbf{s}^{ij} = \boldsymbol{\delta}^{ij}$ and $\mathbf{g}^{\mu\nu} = \boldsymbol{\eta}^{\mu\nu} + \mathbf{h}^{\mu\nu}$; $\boldsymbol{\eta}^{\mu\nu}$ is the Minkowski tensor and $\mathbf{h}^{\mu\nu} = \mathbf{h}^{\mu\nu}(x^\mu)$ the gravitation field tensor of general relativity.

If we now consider the full 10-dimension space-time, we assume that approximately $\mathbf{g}^{\mu\nu}$ and $\mathbf{s}^{ij}$ are not very different from previously but $\mathbf{g}^{\alpha\beta}$ also contains additional off-diagonal terms $\mathbf{a}^{\mu j}$. Let us look for $\mathbf{\Theta}$ special solutions which can be expressed as a product

$$\mathbf{\Theta}(ct, x, y, z, \xi^1, \ldots, \xi^6) = \mathbf{\Psi}(ct, x, y, z)\mathbf{\Phi}(\xi^1, \ldots, \xi^6), \tag{18}$$

With the expressions equ.(17) for $\mathbf{g}$ and equ.(18) for $\Theta$, the equation of evolution becomes

$$[\mathbf{g}^{\mu\nu}\partial_\mu\partial_\nu + \mathbf{g}^{ij}\partial_i\partial_j + \mathbf{a}^{\mu j}\partial_\mu\partial_j]\boldsymbol{\Psi}(ct,x,y,z)\boldsymbol{\Phi}(\xi^1,...,\xi^6)=0 \tag{19}$$

Multiplying equ.(19) by $\boldsymbol{\Phi}^\dagger$ and integrating over the 6 degrees of freedom of $\mathbf{E}_\perp(6)_\mathrm{M}$ gives

$$\mathbf{g}^{\mu\nu}\partial_\mu\partial_\nu\boldsymbol{\Psi} = -\langle\boldsymbol{\Phi}|\mathbf{s}^{ij}\partial_i\partial_j|\boldsymbol{\Phi}\rangle\boldsymbol{\Psi} - \langle\boldsymbol{\Phi}|\mathbf{a}^{\mu j}\partial_j|\boldsymbol{\Phi}\rangle\partial_\mu\boldsymbol{\Psi} \tag{20}$$

with $\dot{\boldsymbol{\Psi}}_\mu = \partial_\mu\boldsymbol{\Psi}$, or

$$\mathbf{g}^{\mu\nu}\partial_\mu\partial_\nu\boldsymbol{\Psi} = \hbar^{-2}\langle\boldsymbol{\Phi}|\mathbf{s}_{ij}\mathbf{p}^i\mathbf{p}^j|\boldsymbol{\Phi}\rangle\boldsymbol{\Psi} - \langle\boldsymbol{\Phi}|\mathbf{a}^{\mu j}\partial_j|\boldsymbol{\Phi}\rangle\dot{\boldsymbol{\Psi}}_\mu \tag{21}$$

i.e.

$$\mathbf{g}^{\mu\nu}\partial_\mu\partial_\nu\boldsymbol{\Psi} = \hbar^{-2} < \mathbf{s}_{ij}\mathbf{p}^i\mathbf{p}^j > \boldsymbol{\Psi} - \langle\boldsymbol{\Phi}|\mathbf{a}^{\mu j}\partial_j|\boldsymbol{\Phi}\rangle\dot{\boldsymbol{\Psi}}_\mu\,. \tag{22}$$

Let us consider the first term on the right hand side of the above equation; in first approximation let us neglect the second term. We get

$$\mathbf{g}^{\mu\nu}\partial_\mu\partial_\nu\boldsymbol{\Psi} = \hbar^{-2} < \mathbf{s}_{ij}\mathbf{p}^i\mathbf{p}^j > \boldsymbol{\Psi} \tag{23}$$

It can be compared with the Klein-Gordon equation in the presence of a gravitation field

$$\mathbf{g}^{\mu\nu}\partial_\mu\partial_\nu\boldsymbol{\Psi}(ct,x,y,z) = (mc/\hbar)^2\boldsymbol{\Psi}(ct,x,y,z) \tag{24}$$

where *m* is the particle mass. Identifying *m* with the self energy term in equ.(23), i.e.

$$mc =< \mathbf{s}_{ij}\mathbf{p}^i\mathbf{p}^j >^{1/2}, \tag{25}$$

means that the particle mass exclusively originates from the 6 extra dimensions of space-time, i.e. from the symmetries of the local orthogonal space $\mathbf{E}_\perp(6)_\mathrm{M}$.

Now let us focus on the second term on the right hand side of equ.(22). It can be interpreted as the coupling between $\boldsymbol{\Psi}$ and an additional vector field

$$\mathbf{A}^\mu = \langle\boldsymbol{\Phi}|\mathbf{a}^{\mu j}\partial_j|\boldsymbol{\Phi}\rangle \tag{26}$$

so that the field equation for $\boldsymbol{\Psi}$ can be written as

$$\mathbf{g}^{\mu\nu}\partial_\mu\partial_\nu\boldsymbol{\Psi} = (mc/\hbar)^2\boldsymbol{\Psi} - \mathbf{A}^\mu\dot{\boldsymbol{\Psi}}_\mu\,. \tag{27}$$

This additional field expresses the connection between gravitation in the “orbital” 4-dimension space-time and the other fundamental interactions which have their origin in the symmetry properties of the "internal" space $\mathbf{E}_{\perp}(6)_{\mathrm{M}}$. That also shows that $\mathbf{a}^{\mu j}$ has only 4 independent components.

The existence of this hypothetical vector companion of gravity would have consequences which can be experimentally tested. The way the gravitation field has been here above introduced in the 4-dimension space-time naturally implies the equivalence between gravitational mass and inertial mass. However if we consider the full $\mathbf{E}(10)$ space we have seen that the hadrons experience a symmetry $\mathbf{G}_{\mathrm{H}}(6)$ isomorphous to $\mathbf{SU}(3)$ whereas the leptons experience a symmetry $\mathbf{G}_{\mathrm{L}}(6)$ isomorphous to $\mathbf{SU}(2)\times\mathbf{U}(1)$. As a consequence, the field $\mathbf{A}^{\mu}$ should act differently upon hadrons and leptons and thus bodies with different compositions would not behave the same way under $\mathbf{A}^{\mu}$. That would result in an apparent violation of the equivalence principle which can be experimentally tested [11] [12].

We note that $\mathbf{A}^{\mu}$ is formally quite similar to the electromagnetic field, the number of hadrons (or anti-hadrons) and the number of leptons (or anti-leptons) *a priori* playing the role of the electric charge. Moreover we see that the form of the 10-dimension metric tensor $\mathbf{g}^{\alpha\beta}$ in equ.(17) presents some obvious similarities with the form of the metric tensor ${\mathbf{g}^{\alpha\beta}}_{KK}$ in the original 5-dimension Kaluza-Klein model in which the $\mathbf{g}^{55}$ component corresponds to a scalar field and the $\mathbf{g}^{\mu 5}$ components correspond to the electromagnetic field, all the more than the 24 off-diagonal components $\mathbf{a}^{\mu j}$ of $\mathbf{g}^{\alpha\beta}$ can be reduced to the 4 components of $\mathbf{A}^{\mu}$:

$$
{\mathbf{g}^{\alpha\beta}}_{KK} = \left( \begin{array}{c|c} \begin{array}{c} \mathbf{g}^{\mu\nu} \\ 4\times 4 \end{array} & \mathbf{g}^{\mu 5} = \mathbf{a}^{\mu} \\ \hline \mathbf{g}^{5\nu} = \mathbf{a}^{\nu} & \mathbf{g}^{55} = \mathbf{s} \end{array} \right) \tag{28}
$$

Finally, in presence of a gravitation field described by the metric tensor $\mathbf{g}^{\mu\nu}$ the dynamics of $\mathbf{A}^{\mu}$ can be accounted for in the 4-dimension space-time by the following effective Lagrangian

$$
\mathfrak{L}_{\mathrm{eff}} = \lambda^{2}\mathbf{g}_{\mu\nu}\mathbf{A}^{\mu}\mathbf{A}^{\nu} + \mathbf{g}_{\mu\nu}\mathbf{g}_{\rho\sigma}\partial_{\rho}\mathbf{A}^{\mu}\partial_{\sigma}\mathbf{A}^{\nu} \tag{29}
$$

It describes an hypothetical new quasi-particle. If $\lambda \neq 0$, it is sensitive to gravitation and its effective mass is $\lambda\hbar / c$; it could then contribute to dark matter via the first term of equ.(29)

and to dark energy via the second term.

## VI. CONCLUSION

We have given a coherent presentation of the fundamental interactions by considering the symmetry properties of a 10-dimension pseudo euclidian space-time. The usual space-time, a 4-dimensional surface $\mathbf{\Sigma}(4)$ whose local symmetry is the Lorentz group $\mathbf{\Lambda}(4)$, is embedded in a flat 10-dimensional space $\mathbf{E}(10)$. In every point of $\mathbf{\Sigma}(4)$ there is a 6-dimension "internal" space $\mathbf{E}_{\perp}(6)_{\mathrm{M}}$ orthogonal to the surface. Simple assumptions about the geometry of $\mathbf{E}_{\perp}(6)_{\mathrm{M}}$: allow to derive the following:

- there are 2 types of particles: (i) hadrons associated to the irreducible representations of $\mathbf{G}_{\mathrm{H}}(6)$ or $\mathbf{SU}(3)$ and experiencing a strong interaction mediated by 8 vector-type gluons, and (ii) leptons associated to the irrducible representations of $\mathbf{G}_{\mathrm{L}}(6)$ or $\mathbf{SU}(2)\times\mathbf{U}(1)$ and experiencing a pseudo electro-weak interaction mediated by 4 vector-type gluinos,

- there are 3 interactions common to both hadrons and leptons and which can be identified with the weak, electromagnetic and Higgs interactions and associated to gauge fields of respective symmetry $\mathbf{SO}(3)$ or $\mathbf{SU}(2)$, $\mathbf{SO}(2)$ or $\mathbf{U}(1)$, $\mathbf{SO}(1)$ or $\mathbf{I}(1)$, and respectively mediated by 3 vector-type bosons ($\mathrm{W}^{+}$, $\mathrm{W}^{-}$, $\mathrm{Z}^{0}$), 1 vector-type photon, plus a scalar boson identified with the Higgs boson.

The hypothetical existence of a pseudo electro-weak interaction mediated by the 4 gluinos is the major difference with the Standard Model based upon the $\mathbf{SU}(3)\times\mathbf{SU}(2)\times\mathbf{U}(1)$ gauge symmetry. For leptons, that pseudo electro-weak interaction is superposed to the usual electro-weak interaction which has the same symmetry.

The particle mass originates from the 6 extra dimensions of space-time, i.e. from the symmetries of the local orthogonal space $\mathbf{E}_{\perp}(6)_{\mathrm{M}}$. When gravitation is introduced, the connection between the "orbital" and the "internal" degrees of freedom evidences an effective additional coupling mediated by a vector-type field. This hypothetical companion of gravity does not act the same way with hadrons and leptons and could be experimentally evidenced through an apparent violation of the equivalence principle. It could also contribute to dark mater and dark energy.

To conclude, pointing out that no compactification of the extra dimensions was needed, let us come back to the question of their size; is it really meaningful ? We experience space and time with our senses or through our instruments but the way we perceive space and time is quite different. For instance, the smallest distance we can perceive with our eyes is of the order of 0.1 mm, and the smallest time interval we can feel is about 0.04 s (the interval between two images in a movie). If we convert those 0.04 s into length (x=ct) we get 12 000 km, 11 orders of magnitude larger than 0.1 mm. In the same manner as space and time are differently perceived and cannot be compared via that this type of conversion, the 6 extra dimensions might be of another nature than usual space or time. Nevertheless we actually feel those extra dimensions inasmuch as “the fundamental interactions of particle physics” can be interpreted as the manifestations in our usual world of the geometry of the extra dimensions. Like the speed of light connects space and time, the connection between those 6 extra dimensions and the 3 usual dimensions of space must involve another physical constant, which should naturally be connected with the gravitation constant.

## ANNEX A

We here consider the subgroups of $\mathbf{SO}(6)$. In the base $\{\xi^i\}$ any infinitesimal transformation of $\mathbf{SO}(6)$ can be written as

$$\mathbf{T}_6 = \mathbf{I}(6) + \mathbf{M} \text{ with } \mathbf{M} = \begin{pmatrix} \mathbf{A}' & \mathbf{B} \\ -\tilde{\mathbf{B}} & \mathbf{A}'' \end{pmatrix} \qquad \text{(A1)}$$

$\mathbf{M}$ is a fully antisymmetric real 6x6 matrix, $\mathbf{A}'$ and $\mathbf{A}''$ are two antisymmetric 3x3 matrices, $\mathbf{B}$ is a 3x3 matrix and $\tilde{\mathbf{B}}$ denotes the transposed matrix of $\mathbf{B}$; $\mathbf{SO}(6)$ has got 15 infinitesimal generators.

$\mathbf{B}$ can be written as $\mathbf{S}+\mathbf{A}+\mathbf{Q}$, where $\mathbf{S}$ is a scalar times the 3-dimension identity $\mathbf{I}(3)$, $\mathbf{A}$ is a 3x3 antisymmetric matrix and $\mathbf{Q}$ a null-trace 3x3 symmetric matrix. $\mathbf{M}$ can then be expressed as the sum of 3 matrices

$$\mathbf{M} = \begin{pmatrix} \mathbf{A}_0 & \mathbf{S} \\ -\mathbf{S} & \mathbf{A}_0 \end{pmatrix} + \begin{pmatrix} \mathbf{A}_2 & \mathbf{Q} \\ -\mathbf{Q} & \mathbf{A}_2 \end{pmatrix} + \begin{pmatrix} \mathbf{A}' - \mathbf{A}_0 - \mathbf{A}_2 & \mathbf{A} \\ \mathbf{A} & \mathbf{A}'' - \mathbf{A}_0 - \mathbf{A}_2 \end{pmatrix} \qquad \text{(A2)}$$

It can be checked from their commutation relations that the matrices $\begin{pmatrix} \mathbf{A}_0 & \mathbf{S} \\ -\mathbf{S} & \mathbf{A}_0 \end{pmatrix}$ and $\begin{pmatrix} \mathbf{A}_2 & \mathbf{Q} \\ -\mathbf{Q} & \mathbf{A}_2 \end{pmatrix}$ generate 2 subgroups $\mathbf{G}_\mathrm{L}(6)$ and $\mathbf{G}_\mathrm{H}(6)$ with respectively 4 and 8 infinitesimal generators; moreover $\mathbf{G}_\mathrm{L}(6)$ has the same Lie algebra as $\mathbf{SU}(2)\times\mathbf{U}(1)$ and that $\mathbf{G}_\mathrm{H}(6)$ has the same Lie algebra as $\mathbf{SU}(3)$. $\mathbf{G}_\mathrm{L}(6)$ and $\mathbf{G}_\mathrm{H}(6)$ have rather similar structures since

$$\begin{pmatrix} \mathbf{A}_0 & \mathbf{S} \\ -\mathbf{S} & \mathbf{A}_0 \end{pmatrix} = \mathbf{A}_0 \otimes \mathbf{I}(2) + \mathbf{S} \otimes \mathbf{J} \qquad \text{(A3)}$$

and

$$\begin{pmatrix} \mathbf{A}_2 & \mathbf{Q} \\ -\mathbf{Q} & \mathbf{A}_2 \end{pmatrix} = \mathbf{A}_2 \otimes \mathbf{I}(2) + \mathbf{Q} \otimes \mathbf{J} \qquad \text{(A4)}$$

where $\mathbf{J} = \begin{pmatrix} 0 & 1 \\ -1 & 0 \end{pmatrix}$ is the single generator of the special orthogonal group $\mathbf{SO}(2)$.

We also note that $\mathbf{G}_{\mathrm{L}}(6)$ and $\mathbf{G}_{\mathrm{H}}(6)$ are invariant, or distinguished, subgroups of $\mathbf{SO}(6)$.

Conversely the matrices $\begin{pmatrix} \mathbf{A}' - \mathbf{A}_0 - \mathbf{A}_2 & \mathbf{A} \\ \mathbf{A} & \mathbf{A}'' - \mathbf{A}_0 - \mathbf{A}_2 \end{pmatrix}$ do not generate an invariant subgroup of $\mathbf{SO}(6)$ since they can be written as

$$\left( \frac{\mathbf{A}' + \mathbf{A}''}{2} - \mathbf{A}_0 - \mathbf{A}_2 \right) \otimes \mathbf{I}(2) + \mathbf{A} \otimes \sigma_{\mathrm{x}} + \left( \frac{\mathbf{A}' - \mathbf{A}''}{2} \right) \otimes \sigma_{\mathrm{z}} \tag{A5}$$

where the $\sigma_{\mathrm{i}}$ are the Pauli matrices, and the pair ($\sigma_{\mathrm{x}}$, $\sigma_{\mathrm{z}}$) does not generate a group.

# ANNEX B

In the main section, we have evidenced one scalar interaction (identified with the Higgs field) and a set of fields respectively mediated through 8 gluons, 4 gluinos, 3 bosons identified with the $W^+$, $W^-$, $Z^0$ and 1 photon, each of them being associated to a symmetry group $\mathbf{G}$. The elementary operations associated to any of them have the form

$$\mathbf{T} = \mathbf{I} + \mathrm{i}\sum_p \mathbf{N}_p \tag{B1}$$

where $\mathbf{N}_p$ is an infinitesimal generator of $\mathbf{G}$. More generally, any operation of $\mathbf{G}$ can be written as

$$\mathbf{T} = \exp \mathrm{i}\sum_p \mathbf{N}_p \tag{B2}$$

We now consider a scalar particle of mass $m$ (but the procedure can be generalized to any spin) associated to the field $\mathbf{\Psi}(ct, x, y, z)$ in the 4-dimension space-time. Its Lagrangian density is

$$\mathfrak{L} = \boldsymbol{\eta}^{\mu\nu}\dot{\mathbf{\Psi}}_\mu{}^\dagger\dot{\mathbf{\Psi}}_\nu + (mc/\hbar)^2\mathbf{\Psi}^\dagger\mathbf{\Psi}\text{, with } \dot{\mathbf{\Psi}}_\mu = \partial_\mu\mathbf{\Psi} \tag{B3}$$

or

$$\mathfrak{L} = \boldsymbol{\eta}_{\mu\nu}\hbar^{-2}[\mathbf{p}^\mu\mathbf{\Psi}]^\dagger[\mathbf{p}^\nu\mathbf{\Psi}] + (mc/\hbar)^2\mathbf{\Psi}^\dagger\mathbf{\Psi} \tag{B4}$$

with $\mu$ and $\nu$ ranging from 0 to 3.

If a global symmetry $\mathbf{T} = \exp \mathrm{i}\sum_p \mathbf{N}_p$ is assumed to be a local one, i.e. $\mathbf{N}_p = \mathbf{N}_p(x^\mu)$, then

$$\mathbf{p}^\mu\mathbf{\Psi} \rightarrow \mathbf{p}^\mu\left(\exp \mathrm{i}\sum_p \mathbf{N}_p\right)\mathbf{\Psi} = \left(\exp \mathrm{i}\sum_p \mathbf{N}_p\right)\left(\mathbf{p}^\mu\mathbf{\Psi} + \hbar\sum_p \partial^\mu\mathbf{N}_p\mathbf{\Psi}\right) \tag{B5}$$

Writing the full state of the particle as a product of the "orbital" state $\mathbf{\Psi}(ct, x, y, z)$ times an "internal" state $\mathbf{\Phi}(\xi^1,...,\xi^6)$, the conservation of the Lagrangian requires the existence of a vector-type gauge field

$$\mathbf{A}^\mu = \sum_p \partial^\mu \left\langle \Phi \middle| \mathbf{N}_p \middle| \Phi \right\rangle \tag{B6}$$

after averaging over the 6 degrees of freedom { $\xi^i$ } since $\mathbf{N}_p$ exclusively acts upon the variables { $\xi^i$ }. It implies the vector character of the above fields: gluons, gluinos, $W^+$, $W^-$, $Z^0$ and photon.

**References**

---